\documentclass[twocolumn,showpacs,preprintnumbers,amsmath,amssymb]{revtex4}

\usepackage{graphicx}
\usepackage{dcolumn}
\usepackage{bm}

\begin{document}

\preprint{APS/123-QED}

\title{High Field ESR in the Two-Dimensional Triangular Lattice Antiferromagnet NiGa$_2$S$_4$}

\author{H. Yamaguchi$^1$, S. Kimura$^1$, M. Hagiwara$^1$, Y. Nambu$^{2, 3}$, S. Nakatsuji$^2$, Y. Maeno$^3$ and 
K. Kindo$^2$}
\affiliation{
$^1$KYOKUGEN, Osaka University, Machikaneyama 1-3, Toyanaka 560-8531, Japan \\ 
$^2$Institute for Solid State Physics, University of Tokyo, 5-1-5 Kashiwanoha, Kashiwa, Chiba 277-8581, Japan \\
$^3$Department of Physics, Kyoto University, Kyoto 606-8502, Japan
}

Second institution and/or address\\
This line break forced

\date{\today}

\begin{abstract}
We report the experimental results of high field ESR measurements in magnetic fields up to about 53 T on the quasi two-dimensional triangular lattice antiferromagnet NiGa$_2$S$_4$. From the temperature evolution of the ESR absorption linewidth, we find three distinct temperature regions, (i) above 23 K, (ii) between 23 K and 8.5 K and (iii) below 8.5 K. The linewidth is affected by the dynamics of $Z_2$-vortices below 23 K and one of conventional spiral resonance modes well explains the frequency dependence of the ESR resonance fields much below 8.5 K. Furthermore, we find an anomaly of the magnetization curve around one third of the saturation magnetization for $H$$\perp$$c$, corresponding to the softening of another resonance mode. These results suggest an occurrence of $Z_2$ vortex-induced topological transition at 8.5 K. 
\end{abstract}

\maketitle
Two-dimensional (2D) geometric frustrated systems have attracted considerable interests over the three decades because they are expected to have a novel ground state such as a spin liquid. Among them, the 2D triangular lattice antiferromagnet (TLAFM) has been studied extensively both theoretically and experimentally. Since a spin liquid state was theoretically suggested in the $S$=1/2 Heisenberg TLAFM in 1973~\cite{anderson}, a lot of attention has been paid to the TLAFM till today. Recent theories, however, proposed an ordered ground state with a 120$^{\circ}$ spin structure in the $S$=1/2 TLAFM with only the nearest-neighbor exchange interactions~\cite{120_1} and a spin liquid in the $S$=1/2 TLAFM with distant neighbor and multiple-spin exchange interactions~\cite{sp_1,sp_2}. Experimentally, there are only a few samples which are highly 2D and show no long range order (LRO) down to extremely low temperatures~\cite{bedtttf}. Therefore, it was difficult to clarify what kind of state is realized actually as a ground state of such a strongly frustrated system, and ideal compounds having highly 2D characters and isotropic interactions have been strongly desired to investigate the ground state of them. Recently, a new quasi 2D $S$=1 TLAFM NiGa$_2$S$_4$ has attracted substantial interests as a candidate with the spin liquid state~\cite{science}. This compound has slabs consisting of two GaS layers and one NiS$_2$ layer which stack along the $c$-axis, and each slab is separated by a van der Waals gap, resulting in highly 2D character. In NiS$_2$ layer, Ni$^{2+}$ (3$d$$^8$, $S$=1) ions form an exact triangular lattice. Despite the fact that strong AF interactions are suggested from a large Weiss temperature ${\theta}\rm{_w}$$\simeq$$-80$ K, it has no LRO down to 0.08 K~\cite{NMR}, and some noteworthy characteristics were observed~\cite{science}. The neutron diffraction measurements have clarified that the AF correlation with a wavevector ${\boldsymbol{q}}$ = (0.158, 0.158, 0) extends to only about 2.5 nm (about seven lattice spacing) even at 1.5 K, which corresponds to a short-range order (SRO) with about 57$^{\circ}$ spiral spin structure.  The magnetic susceptibility at 0.01 T exhibits a weak peak at about 8.5 K ($T^*$) and a finite value at the low-$T$ limit, indicating the existence of gapless excitations. The magnetic specific heat shows broad peaks at about 10 K and 80 K, and $T$$^2$-dependence with no field dependence up to 7 T below 10 K. The $T$$^2$-dependent specific heat indicates the existence of the $k$-linear excitation similar to the Nambu-Goldstone mode associated with a breaking of a continuous symmetry. The $T$$^2$-dependent specific heat was also reported in some 2D frustrated spin systems. These curious experimental results have stimulated further theoretical and experimental studies on this compound in recent years~\cite{NMR,AFQ,jikariron,kawamuraNiGaS,nambusan,takubo,fePRL,single}. In some theoretical works, the  blinear-biquadratic exchange interactions were considered to account for such peculiar experimental results~\cite{AFQ,jikariron,kawamuraNiGaS}. Very recent NMR, NQR and $\mu$SR measurements evidenced the occurrence of a novel state where the spins have a MHz scale dynamics through a weak freezing below about 10 K ~\cite{NMR}.

In this Letter, we report the results of ESR measurements in high magnetic fields up to about 53 T on single crystals of NiGa$_2$S$_4$ to clarify the spin dynamics in more detail. We have found that the dynamics of $Z_2$ vortices affects the temperature dependence of the ESR absorption linewidth and the frequency dependence of the ESR resonance fields at 1.3 K below $T^{\ast}$ is well explained by a conventional spin wave theory. These results suggest an occurrence of $Z_2$ vortex-induced topological transition.  
\begin{figure}[t]
\begin{center}
\includegraphics[width=16pc]{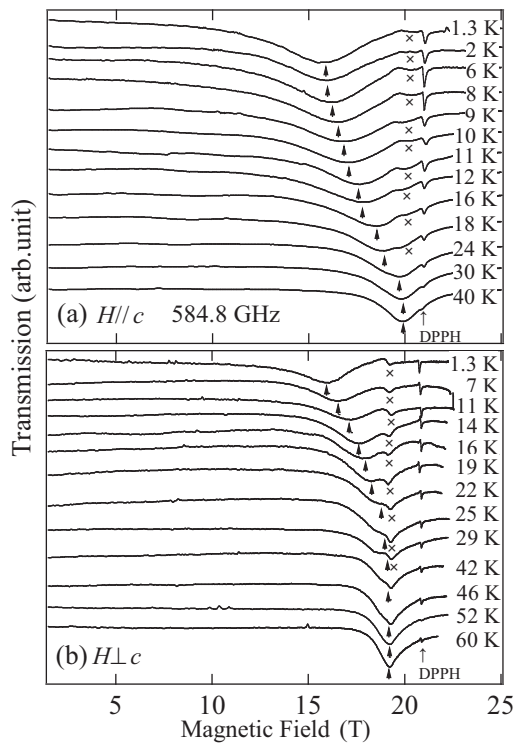}
\caption{Temperature dependence of ESR absorption spectra of NiGa$_2$S$_4$ at 584.8 GHz for (a) $H$$\parallel$$c$ and for (b) $H$$\perp$$c$. The arrow and cross indicate the broad large signal and the weak one, respectively. The sharp signal at about 21 T comes from an ESR standard sample of DPPH for correction of the magnetic field.}\label{f1}
\end{center}
\end{figure}
\begin{figure}[t]
\begin{center}
\includegraphics[width=17pc]{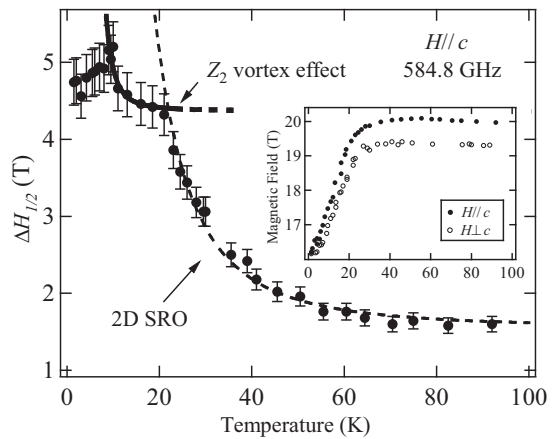}
\caption{Temperature dependence of full width at half maximum of the ESR linewidth in NiGa$_2$S$_4$ for $H$$\parallel$$c$ at 584.8 GHz. Solid and broken lines indicate calculated line widths caused by the dynamics of $Z_2$ vortices and the development of 2D SRO, respectively. The inset shows the temperature dependence of the resonance field of the large ESR signal in NiGa$_2$S$_4$ at 584.8 GHz. Open and closed circles represent the resonance fields for $H$$\parallel$$c$ and $H$$\perp$$c$, respectively.}\label{f2}
\end{center}
\end{figure}

The ESR measurements for the frequencies up to 1.4 THz were conducted in pulsed magnetic fields up to about 53 T at temperatures between 1.3 K and about 90 K. The ESR measurements below 500 GHz in steady magnetic fields up to 14 T at 1.4 K were done by utilizing a superconducting magnet and a vector network analyzer. Single crystals used for the measurements were grown by a chemical-vapour transport technique using iodine~\cite{single}.

Figures 1(a) and 1(b) show the temperature dependence of the ESR absorption spectra at 584.8 GHz in pulsed magnetic fields for $H$$\parallel$$c$ and for $H$$\perp$$c$, respectively. We observed a strong resonance signal indicated by an arrow whose linewidth is extremely broad and the other weak one indicated by a cross. The weak resonance signals are considered to come from the paramagnetic resonances caused by imperfection of the lattices. The temperature dependences of the absorption linewidth and the resonance field of the broad signal are plotted in Fig. 2 and its inset, respectively. The broad ESR signal shifts toward lower field with decreasing temperature for both directions below about 30 K. The absorption linewidth increases as the temperature is lowered from about 80 K to 8.5 K, and a bending appears at 23 K at which the increasing tendency of the linewidth with decreasing temperature becomes gradual. This behavior between 80 K and 23 K is interpreted as the line broadening due to the development of the SRO as suggested from the NQR measurements~\cite{NMR}. 
Thus, we separately examine the temperature ranges above 8.5 K, namely above 23 K and between 23 K and 8.5 K. First, the linewidth above 23 K is investigated in terms of the critical behavior in the 2D Heisenberg AFM. The common expression of temperature dependence of the linewidth (full width at half maximum) associated with the AF LRO temperature $T_{\rm{N}}$ is
\begin{align}
\Delta H_{1/2} \propto   (T-T_{\rm{N}})^{-p},\label{p}
\end{align} 
where $p$ is a critical exponent, which reflects the anisotropy and the dimensionality of the system. Based on Kawasaki's dynamic equations, the critical exponent of 2D AFM is given by $p=(3-2\eta )\nu$, where $\nu$ is the critical exponent for the inverse correlation length and $\eta $ the critical exponent related to the static correlations~\cite{p-riron}. In the case of 2D Heisenberg AFM, $p$ is evaluated to be about 2.5 from the experimentally obtained $\eta=0.2$ and $\nu=0.95$~\cite{p-bunken}. In NiGa$_2$S$_4$, one expects that the development of the SRO gives the same critical behavior as that of the 2D Heisenberg AFM. In addition, since there is no LRO down to 0.08 K~\cite{NMR}, $T_{\rm{N}}$ must be small enough to approximate 0. Thus, we fit Eq. (1) with $p=2.5$ and $T_{\rm{N}}\simeq 0$ to the experimental line width below 80 K and obtained good agreement above 23 K. Below 23 K, the increase of the linewidth with decreasing temperature becomes gradual. This indicates that some effect disturbs the development of the SRO below 23 K. We take into account a topologically stable point defect which is the so called $Z_2$ vortex, characterized by a two-valued topological quantum number, suggested theoretically~\cite{Z2kawamura} in the Heisenberg TLAFM.  Unlike a conventional vortex of spins in the $XY$ model, it is a vortex of spin chiralities. The $Z_2$ vortices form bound pairs at low temperatures and begin to dissociate at a certain critical temperature $T_{\rm{V}}$, causing a Kosterlitz-Thouless-type phase transition. Above $T_{\rm{V}}$ the unbound free vortices are thermally excited, and the spin fluctuates through the passage of the vortices. Hence, unbound $Z_2$ vortices disturb the development of the SRO. The effect of $Z_2$ vortex on the ESR linewidth is discussed for Heisenberg TLAFMs, HCrO$_2$ and LiCrO$_2$~\cite{Z2linewidth}. According to the analysis used for these compounds, the ESR linewidth above $T_{\rm{V}}$ is expected to behave as,       
\begin{align}
\Delta H_{1/2} \propto \exp(\frac{E_{\rm{V}}}{k_{\rm{B}}T}),
\end{align}
where $E_{\rm{V}}$ is an activation energy of a free $Z_2$ vortex~\cite{Z2linewidth} and $k_{\rm{B}}$ the Boltzmann constant. According to the estimation by a Monte Carlo simulation~\cite{Z2kawamura},  $T_{\rm{V}}$ and $E_{\rm{V}}$ are given by $T_{\rm{V}}=0.31JS^2$ and $E_{\rm{V}}=1.65JS^2$, where $J$ is the nearest neighbor exchange interaction, and thus the ratio $E_{\rm{V}}/T_{\rm{V}}$ becomes constant. We take $T_{\rm{V}}=8.5$ K from the anomaly of the magnetic susceptibility, and then obtain $E_{\rm{V}}=45.2$ K. Using this value, satisfactory agreement between experiment and calculation is attained between 8.5 K and 23 K as shown in Fig. 2. From the neutron scattering experiment, it was found that the correlation length in the triangular plane grows up to be about three lattice units at 23 K~\cite{length}. This result indicates that the correlation length of about three lattice units is needed to observe the $Z_2$ vortex effect. The temperature dependent shift of the resonance field below about 30 K in the inset of Fig. 2 is also thought to be caused by growing up of the correlation length which is about two lattice units at 30 K~\cite{length}. 
Below about 10 K, the linewidth shows a gradual decrease as the temperature is lowered and even at the lowest temperature, it remains a large value. From the NQR measurements, Takeya $et$ $al$. suggested that a gradual spin freezing occurs between 10 K and 2 K, and inhomogeneous internal fields appears below 2 K~\cite{NMR}. This considerably large ESR linewidth at the lowest temperature must reflect the distribution of the internal fields. Accordingly, the temperature dependence of the ESR linewidth is consistent with the NQR results.   
\begin{figure}[t]
\begin{center}
\includegraphics[width=7cm]{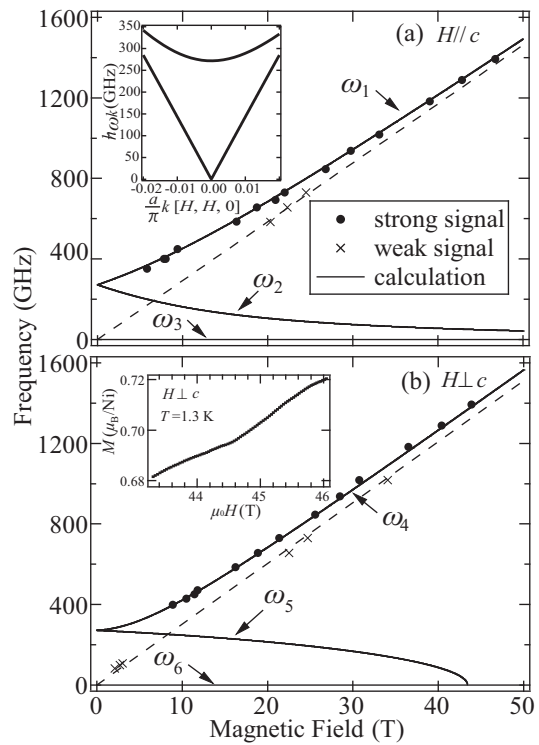}
\caption{Frequency-field diagram of NiGa$_2$S$_4$ at 1.3 K in pulsed magnetic fields and at 1.4 K in steady magnetic fields for (a) $H$$\parallel$$c$ and (b) $H$$\perp$$c$. Closed circles and crosses denote strong and week signals, respectively. Solid and broken lines represent calculated resonance modes and paramagnetic resonance line, respectively. The insets in Figs. 3(a) and 3(b) show the calculated spin wave dispersion relation near $k$=0 at $H$=0 T where $a$ is the lattice parameter and the magnetization curve for $H$$\perp$$c$ at 1.3 K near the softening field around 44 T, respectively.}\label{f3}
\end{center}
\end {figure}

Next, we discuss the frequency dependence of the resonance field of the strong ESR signal at the lowest temperature. Figures 3(a) and 3(b) show the resonance fields at various frequencies obtained from the ESR absorption spectra at 1.3 K in pulsed magnetic fields and at 1.4 K in steady magnetic fields for $H$$\parallel$$c$ and for $H$$\perp$$c$, respectively. The broken line in each figure represents a paramagnetic resonance line with $g$-factor measured at about 90 K. As mentioned before, the weak signal probably comes from isolated Ni ions due to the imperfection of the lattices and puts on the paramagnetic resonance line. The ESR modes of the strong signal have a zero-field gap and approach the paramagnetic resonance modes with increasing fields for both field directions. 
These ESR modes are reminiscent of spiral spin resonance mode observed in TLAFM with easy-plane anisotropy~\cite{spiralmode1}. It is known that even in the disordered phase, some low-dimensional magnets exhibit resonance modes expected in the LRO phase in the temperature region where the SRO well develops~\cite{TMMC}. Therefore, we analyze the experimental results in terms of spiral spin resonances. A conventional mean-field approximation theory derives that the exchange constants are required to be $J$$_1$/$J$$_3$=$-0.2$ and $J$$_2$=0, where $J$$_1${\textless}0 (ferromagnetic) and $J$$_3${\textgreater}0 (antiferromagnetic), to realize the 57$^{\circ}$ spiral spin structure indicated by the neutron diffraction measurements. The $J$$_i$ ($i$ = 1, 2, 3) represent the first, second, and third neighbor exchange interactions, respectively. The existence of considerably strong third neighbor exchange interaction is also suggested in Ref.~\cite{takubo}. Since $J$$_3$ is large enough compared to the others, we consider $J$$_1$=0 for simplicity.  
From the conventional spin wave theory, three kinds of excitation modes are extracted for the spiral magnet. In the case of the Heisenberg system with no anisotropy, these modes are degenerate and gapless at zero field. However, when the system has the easy-plane anisotropy ($D${\textgreater}0), two modes have a gap whereas one mode remains gapless. These two excitations with a gap are expected to be observed in the ESR experiment. The resonance modes with 120$^{\circ}$ spiral spin structure for $H$$\parallel$$c$ and $H$$\perp$$c$ were derived by the conventional spin wave theory~\cite{cooper} and molecular field approximation~\cite{120ESR}, respectively. The calculated three resonance modes, namely two gapped and one gapless modes, are drawn in Fig. 3. 
The agreement between experiment and calculation is good by using the following parameter values; $J$$_3$/$k\rm{_B}$=21 K, $D$/$k\rm{_B}$=0.9 K, $g$$_{\parallel}$=2.09, $g$$_{\perp }$=2.15. The zero-field energy gap is evaluated to be about 13 K. From the obtained  $J_3$, we estimate the Weiss temperature to be $-84$ K, which is nearly equal to the value obtained from the magnetic susceptibility measurement~\cite{science}. We did not observe the resonance modes, ${\hslash}$${\omega}$$_2$ and  ${\hslash}$${\omega}$$_5$, which gradually drop with increasing field. These modes have weak field dependence and are expected to show much broader resonance linewidth than the upper ones. Hence, these resonance modes were hardly observed. According to Ref.~\cite{120ESR}, the magnetic field at which the ${\hslash}$${\omega}$$_5$ becomes zero corresponds to the occurrence field of the collinear spin configuration, in which spins on two sublattices are parallel and those on one sublattice are antiparallel to the applied magnetic field in the basal plane. Correspondingly, we observed an anomaly of the magnetization curve for $H$$\perp$$c$ around 44 T as shown in the inset of Fig. 3(b), the magnitude of which is about one third of the saturation magnetization (2.15 $\mu_{\rm B}$/Ni), while no anomaly was observed in the magnetization curve for $H$$\parallel$$c$. 

The above analysis suggests that spin-wave like excitation modes develop at low temperatures in NiGa$_2$S$_4$. Importantly, the lowest gapless mode derived from our analysis has $k$-linear dispersion at $k$ ${\sim}$ 0 as shown in the inset of Fig. 3(a). This mode remains gapless even in finite magnetic fields. Furthermore, spin wave velocity $v_{\rm{s}}$, which corresponds to a slope of the $k$-linear dispersion, is almost constant in the field low enough compared with the saturation field, which is evaluated to be about 130 T. Actually, $v_{\rm{s}}=948$ m/s at 7 T calculated for $H$ $\parallel$ $c$ with the parameters determined from the ESR measurements is almost the same as $v_{\rm{s}}=949$ m/s at 0 T. In addition, it is considered that the gapped modes does not affect the specific heat below 10 K because of the energy gap evaluated to be about 13 K. These features can qualitatively explain the field independent specific heat with $T^2$-dependence. From the magnon with $k$-linear dispersion on the 2D triangular lattice, we obtain the specific heat at low temperature region as $C=(3{\zeta}(3)/{\pi})N{\rm{_A}}k_{\rm{B}}^3(T/v_{\rm{s}})^2$, where $N\rm{_A}$ is the Avogadro number and ${\zeta}(3) = 1.202$. The calculated result, however, is somewhat smaller than the experimental one. For a quantitative agreement with the experimental specific heat, the value of $v_{\rm{s}}$ is required to be nearly three times smaller than that expected from our ESR analysis. From the conventional spin wave theory, a well-defined sharp energy dispersion is expected. On the other hand, the distribution of the internal field in NiGa$_2$S$_4$ likely gives broad energy structure of the excitation modes, which were implied from the considerably large ESR linewidth. This broadening of the energy spectrum may cause specific heat larger than that expected from the spin wave theory. For direct observation of the energy dispersion, detailed inelastic neutron scattering experiment is desired.

Finally, we discuss the possibility of the $Z_2$ vortex scenario in Ref.~\cite{kawamuraNiGaS}. If the $Z_2$ vortex-induced topological transition with small biquadratic exchange interaction occurs at $T_{\rm{v}}$, the low-temperature phase is dominated by spin wave excitations, and the spin correlation length remains finite~\cite{kawamuraNiGaS}. In our observation, below $T_{\rm{v}}$${\simeq}$8.5 K, the experimental results suggest spin wave like excitations. In addition, the $Z_2$ vortices effect can be found in the temperature dependence of the ESR absorption linewidth. Consequently, the unconventional properties of NiGa$_2$S$_4$ in the low temperature region strongly suggests such a vortex-induced topological transition.

This work was partly supported  by the Grants-in-Aid for Scientific Research on Priority Areas (No. 17072005 and No. 19052003) and Scientific Research (B) (No. 20340089) and by 21st Century COE Program named "Towards a new basic science: depth and synthesis" and Global COE Program named "Core Research and 
Engineering of Advanced Materials-Interdisciplinary Education Center for 
Materials Science", all from the Ministry of Education, Science, Sports, Culture and Technology in Japan.


\end{document}